\documentclass[dvips]{article}		
\usepackage{subfigure}
\usepackage{icrctc07}

\title{Photocathode-Uniformity Tests of the Hamamatsu R5912 Photomultiplier Tubes Used in the Milagro Experiment}
\shorttitle{Tests on the R5912 Hamamatsu PMT}

\authors{
V.~Vasileiou$^{1}$, R. W. Ellsworth$^2$, A. Smith$^1$
}
\shortauthors{V. Vasileiou et al. }

\afiliations{
$^1$Department of Physics, University of Maryland, College Park, MD \\
$^2$Department of Physics and Astronomy, George Mason University, Fairfax, VA
}

\abstract{
The Milagro experiment observes the extensive air showers produced
by very high energy $\gamma$-rays impacting the Earth's atmosphere. Milagro
uses 898 Hamamatsu R5912 Photomultiplier Tubes. To complete our Monte
Carlo simulations, we tested the photocathode uniformity of our PMTs.
The main finding was that the PMT gain and detection efficiency
are a function of the distance from the center of the photocathode.
Both quantities become considerably smaller as the illumination position
nears the edge of the photocathode.
}

\email{vlasisva@umd.edu}
\begin{document}
\maketitle

\section{Introduction}

Milagro \cite{milagro} is a water-Cherenkov detector
at an elevation of 2650m at the Jemez Mountains in New Mexico. It
comprises a central 60m x 80m x 8m pond surrounded by a 200m
x 200m array of 175 ``outrigger'' tanks. The pond, covered with
a light barrier, is instrumented with two layers of photomultiplier
tubes (PMTs). The top ``air-shower'' layer consists of 450 PMTs 
while the bottom ``muon'' layer has 273 PMTs.
Each outrigger tank contains $\sim$4000l of water and one PMT. The PMTs collect the Cherenkov
light produced by the air shower particles, as they transverse the
detector's water volume. The AS and OR layers allow the accurate
measurement of the air shower particle arrival times used for reconstructing the 
direction of the shower-initiating particle.
The greater depth of the muon layer ($\sim$17 radiation
lengths) is used to distinguish deeply penetrating muons and hadrons,
which are common in hadron induced air showers, from electrons and $\gamma$-rays.

The motivation for the uniformity tests described here was
a disagreement between the single-muon response of the muon-layer PMTs
and a Monte Carlo simulation of this response. According to the PMT tests performed
by the IceCube experiment\footnote{http://www.ppl.phys.chiba-u.jp/research/IceCube/docs/presentations/index.html},
the photocathode response of their 10' PMTs is not uniform all 
over the photocathode's surface. Before the experimental work reported in this paper,
the Milagro PMTs were simulated as having the same properties all over the face of the PMT. 
We thought it possible that the way the PMTs were simulated was the cause of the disagreement 
regarding the single-muon response of the muon-layer PMTs.

For that reason, an apparatus was constructed to examine the dependence of the gain
and detection efficiency on the distance of the photon-detection position
from the center of the photocathode. 

\section{\label{sec:Experimental-setup}Experimental setup}

A light source that produced a narrow parallel beam of light
was constructed. Its body was composed
of two threaded tubes and a ND2 optical filter between them.
Two end-caps were attached to the edges of the light source.
Each end-cap had a hole of 4.8mm diameter. One of the end-caps was
used to firmly hold an LED and the other as a collimator. The light
was produced by a 5mm diameter red NSPR518AS LED by Nichia Corp. 
Its biggest advantage is that its spectral response
is narrow ($<$30nm FWHM) and that it is stable under changes in temperature.
The resulting light beam had an opening angle of $\sim1^{o}$ . 

Three Hamamatsu R5912 PMTs from Milagro were tested. From these, one
(PMT \#1024) was extensively tested. 

The tests required the light source to be positioned in predefined
positions with respect to the PMT face. A wooden
structure was built for that purpose.
With the help of clamps, the PMT and the light source were mounted
steadily on the structure. The light source was always touching the
surface of the glass and was oriented perpendicularly to it.

In all the tests, the PMT was in an optically shielded black box.
Before each test, the PMT was conditioned under voltage until its
dark noise rate fell to normal levels (\textless{}2KHz).
The magnetic field at the location
of the tests was measured to be about 0.5Gauss. Unless otherwise
noted, the tests were made using PMT \#1024, the supply voltage was
1800V and the PMT was in the vertical position (photocathode facing
up).

\section{The tests}

\subsection{Examining the gain of the PMT}

The best way to examine the gain was to measure the charge of the
pulses. However, due to the fact that no instruments with this capability
were available, we measured the pulse heights instead. As it was found,
the widths of the pulses were almost independent of their height.
For that reason, we could assume that the charge of a pulse was directly
proportional to its height. Thus, the gain could be examined
by measuring the pulse heights.

For making the pulse height distributions, the PMT pulses were analyzed
with a digital oscilloscope. For this test, the light source was operated in pulsed mode. In this
mode a pulse generator was used
for driving the light source and for triggering the oscilloscope.
The oscilloscope measured the amplitudes of the PMT pulses,
and sent the results to a PC through the serial port. 
The light level was low enough that the PMT detected a signal from
only a small percentage of the light pulses, so almost all of the PMT pulses
corresponded to a single PE. The dark rate of the PMT was about 2KHz, so on average there was just one dark noise
hit digitized every $\sim2\cdot10^{4}$ triggers. 

In all the following pulse height spectra, the pedestal
\footnote{Pulse height\textless{}5mV} is suppressed and the curves are normalized.

\subsubsection{Dependence of the gain on the photon-detection position}

Pulse height distributions were made for different illumination positions
on the photocathode.  One of the PMTs was extensively
tested (PMT \#1024). For the other two PMTs, only the side (equator)
and top illumination positions were examined.

\begin{figure}[!ht]
\begin{centering}
\includegraphics[bb=10bp 0bp 520bp 350bp,clip,width=1\columnwidth]{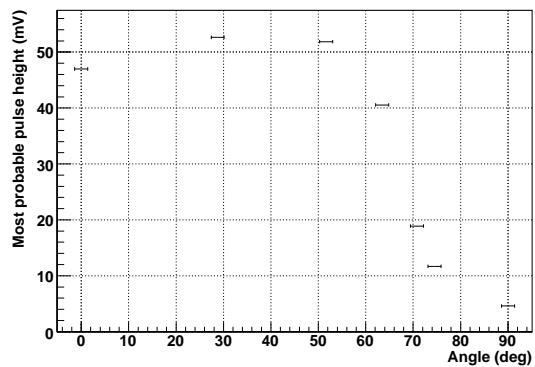}
\par\end{centering}

\caption{\label{cap:Most-probable-pulse}Most probable pulse height for each
of the examined illumination positions}
\end{figure}

\begin{figure}[!ht]
\begin{centering}
\includegraphics[bb=25bp 70bp 374bp 737bp,clip,width=1\columnwidth]{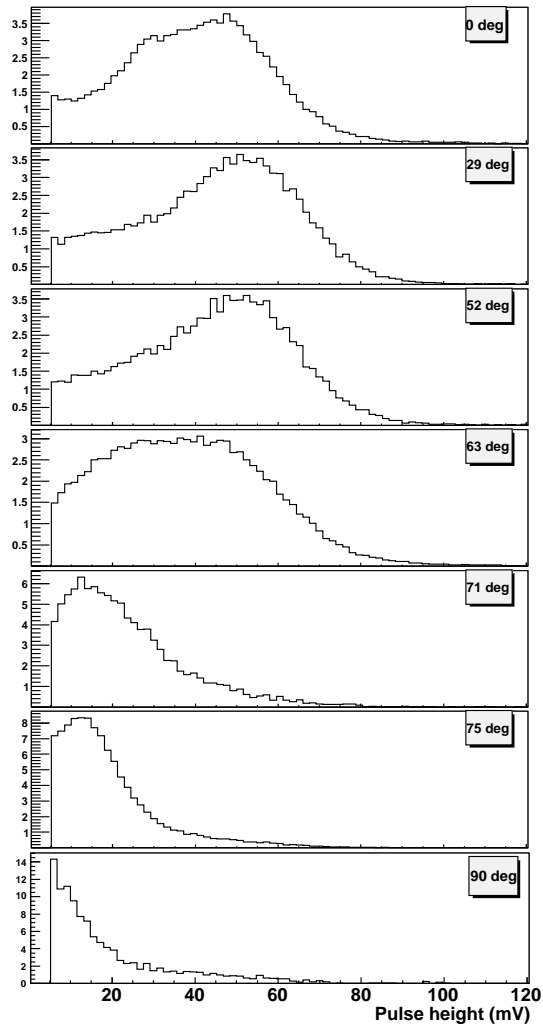}
\par\end{centering}

\caption{\label{cap:Pulse-height-distributionfor diffpos}Pulse height distribution
for illumination of different positions on the photocathode}
\end{figure}

The results of the tests on PMT \#1024 are shown in figures \ref{cap:Most-probable-pulse} and 
\ref{cap:Pulse-height-distributionfor diffpos}. The
angle of an illumination position is defined as the distance
of this position from the center of the photocathode (measured
on the surface of the PMT) over the total distance from the center to the equator
times $90^{o}.$ This way, $0^{o}$ corresponds to the center of the
photocathode and $90^{o}$ to the equator.
The gain was significantly reduced when the illumination position was near the the equator of the PMT.
While it remained almost constant from the center of the photocathode to about
$52^{o}$, there was a sharp transition after that point, with the
gain reducing in size by a factor of $\sim2.7$ in only $20^{o}$. 

The gain for illumination at $90^{o}$ was
also significantly reduced in the other two tested PMTs. 

\subsubsection{Dependence of the gain on the supply voltage and PMT orientation
for illumination at the top and the side of the photocathode}

To further examine the dependence of the gain on the photon-detection
position, we tried to find any correlations between this dependence
and the supply voltage and PMT orientation. Pulse height distributions
were made for different supply voltages (1600V, 1800V and 2000V),
illumination positions (top and near the equator), and PMT orientations
(horizontal and vertical). While in the horizontal orientation, a
vector which was perpendicular to the center of the photocathode and pointing outwards,
pointed south. 

\begin{figure}[!ht]
\begin{centering}
\includegraphics[bb=10bp 0bp 520bp 355bp,clip,width=1\columnwidth,keepaspectratio]{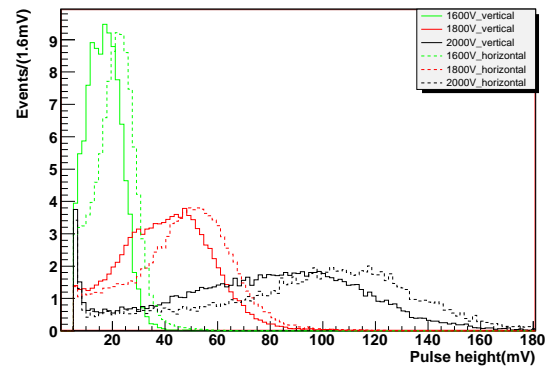}
\par\end{centering}

\caption{\label{fig:PHA_TOP_SUPPLY_ORIENTATION}Pulse height distribution
for illumination at the center of the photocathode for different PMT
orientations and supply voltages.}
\end{figure}

\begin{figure}[!ht]
\begin{centering}
\includegraphics[bb=10bp 0bp 520bp 355bp,clip,width=1\columnwidth,keepaspectratio]{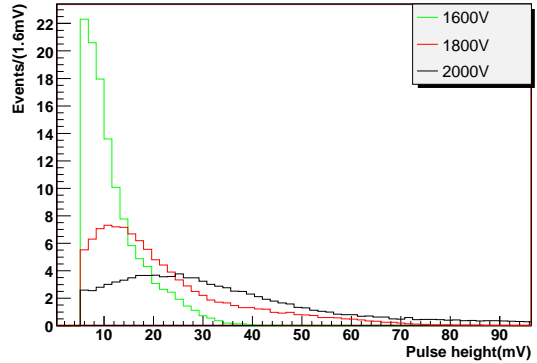}
\par\end{centering}

\caption{\label{fig:PHA_HORIZONTAL_VOLTAGES}Pulse height distribution for
illumination of a point near the equator with the PMT horizontal for
different supply voltages.}
\end{figure}

\begin{figure}[!ht]
\begin{centering}
\includegraphics[bb=10bp 0bp 520bp 355bp,clip,width=1\columnwidth,keepaspectratio]{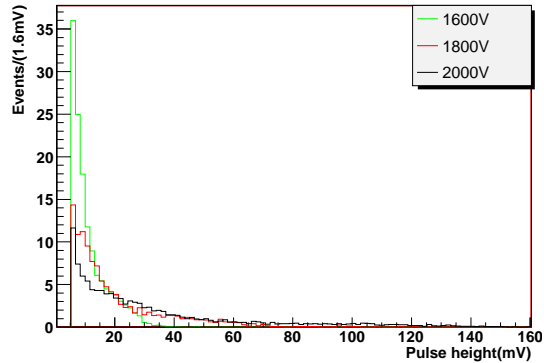}
\par\end{centering}

\caption{\label{fig:PHA_VERTICAL_VOLTAGES}Pulse height distribution for illumination
of a point near the equator with the PMT vertical for three different
supply voltages.}
\end{figure}

Initially, the gain was examined with illumination of the center of the
photocathode (figure \ref{fig:PHA_TOP_SUPPLY_ORIENTATION}). Only a small
dependence of the gain on the PMT orientation was found. Next,
the same PMT orientations and supply voltages were tested for
illumination of the side of the photocathode. As in the previous section,
the gain for illumination at the side of the PMT was considerably
lower than the one for illumination at the top (figures \ref{fig:PHA_HORIZONTAL_VOLTAGES}
and \ref{fig:PHA_VERTICAL_VOLTAGES}). Again, in the horizontal position
the gain was slightly higher than in the vertical position. Increasing
the supply voltage (and as a consequence, the voltage between the
photocathode and the first dynode) increased the gain only when the
PMT was in the horizontal position.


\subsection{Dependence of the detection efficiency on the photon-detection position}

The light source was connected to a voltage supply that was always on. The PMT signal was sent to
a discriminator, which was connected to a scaler. A gate generator,
used as a timer, was used to start and stop the discriminator. When
the gate generator was on, the discriminator compared the pulse heights
against its lowest discrimination threshold (10mV), and the scaler
counted the triggers from the discriminator. The measurements were multiplied by an 
appropriate counting-efficiency factor to account for the pulses 
between the pedestal (5mV) and the discrimination threshold (10mV) that could not
be counted.

There were seven different sets of measurements, each one corresponding to
a different illumination position.
Each set was composed of two background measurements, followed by two signal
measurements and then followed by two background measurements. The
duration of each measurement was 10sec. The count rate, when the
PMT was illuminated, was of the order of tens of KHz. 

Table \ref{cap:Relative-detection-efficiency} shows the 
relative detection efficiency for illumination at various points
of the photocathode vs illumination at its center.

\begin{center}

\begin{table}[!ht]
\begin{centering}
\begin{tabular}{|c|c|c|c|}
\hline 
{\scriptsize Angle (deg)}&
\multicolumn{3}{c|}{{\scriptsize Relative detection efficiency (\%)}}\tabularnewline
\hline
\hline 
&
\emph{\scriptsize PMT \#1024}&
\emph{\scriptsize PMT \#394}&
\emph{\scriptsize PMT \#992}\tabularnewline
\hline 
{\scriptsize 0}&
{\scriptsize 100$\pm0.5$}&
{\scriptsize 100$\pm0.5$}&
{\scriptsize 100$\pm0.8$}\tabularnewline
\hline 
{\scriptsize 29}&
{\scriptsize 92.4$\pm0.5$}&
&
\tabularnewline
\hline 
{\scriptsize 52}&
{\scriptsize 88.7$\pm0.3$}&
&
\tabularnewline
\hline 
{\scriptsize 63}&
{\scriptsize 62.7$\pm0.3$}&
&
\tabularnewline
\hline 
{\scriptsize 71}&
{\scriptsize 65.5$\pm0.6$}&
&
\tabularnewline
\hline 
{\scriptsize 75}&
{\scriptsize 65.9$\pm0.3$}&
&
\tabularnewline
\hline 
{\scriptsize 90}&
{\scriptsize 44.1$\pm0.2$}&
{\scriptsize 30.8$\pm0.1$}&
{\scriptsize 33.4$\pm0.1$}\tabularnewline
\hline
\end{tabular}
\par\end{centering}

\caption{\label{cap:Relative-detection-efficiency}Relative detection efficiency
vs illumination position}
\end{table}

\par\end{center}

\section{Conclusion}
The photocathode uniformity of the Hamamatsu R5912 PMTs used in the
Milagro experiment was examined. The tests showed that the PMT gain
and detection efficiency are a function of the photon-detection position.
Both quantities are smaller when points near the edge of the photocathode
are illuminated. This effect was present in all tested PMTs and was
not negligible in magnitude.

Before this study, the Milagro PMTs were simulated as having the same
properties all over the face of the PMT. After using the experimental 
results in the simulation of the PMTs, the simulation now predicts 
that an atmospheric muon produces $\sim$110PEs (down from $\sim$200PEs) 
in the PMTs of the muon layer compared with $\sim$100PEs at the experiment.


\begin{thebibliography}{1}

\bibitem{milagro}
{Atkins, R.W.} et~al.
\newblock Tev gamma-ray survey of the northern hemisphere sky using the milagro
  observatory.
\newblock {\em ApJ}, 608:680--685, 2004.

\end{thebibliography}
\end{document}